\documentclass{article}

\usepackage[english]{babel}

\usepackage[letterpaper,top=2cm,bottom=2cm,left=3cm,right=3cm,marginparwidth=1.75cm]{geometry}

\usepackage{harvard}
\usepackage{amsmath}
\usepackage{amssymb}
\usepackage{graphicx}
\usepackage[colorlinks=true, allcolors=blue]{hyperref}
\usepackage{booktabs}

\title{Complex Wavelet-Based Sinogram Segmentation for Metal Artifact Reduction in Cone-Beam CT}

\author{
Siiri Rautio*$^{1}$,
Alexander Meaney$^{1}$,
Salla-Maaria Latva-Äijö$^{1}$,
Harshit Agrawal$^{2,3}$,\\
Mikael Brix$^{4,5}$,
Dinidu Jayakody$^{4}$,
and Samuli Siltanen$^{1}$\\[2mm]
{\small $^{1}$Department of Mathematics and Statistics, University of Helsinki, Finland}\\
{\small$^{2}$Department of Electrical Engineering and Automation, Aalto University School of Electrical Engineering, Finland}\\
{\small$^{3}$Planmeca Oy, Finland}\\
{\small$^{4}$Research Unit of Health Sciences and Technology, University of Oulu, Finland}\\
{\small$^{5}$Department of Diagnostic Radiology, Oulu University Hospital, Finland}
}

\date{}

\begin{document}
\maketitle
{\let\thefootnote\relax\footnote{*Corresponding author, siiri.rautio@helsinki.fi}}
\begin{abstract}

\emph{Objective.} Metal artifacts in cone-beam computed tomography (CBCT) arise from inconsistent projections caused by highly attenuating materials, leading to severe streaking and shading that degrade image quality and hinder clinical interpretation. This work aims to develop a robust, non-learned projection-domain method for metal artifact reduction based on analytical segmentation directly in the three-dimensional sinogram.
\emph{Approach.} We propose a projection-domain metal artifact reduction method that performs metal segmentation in the 3D sinogram using the three-dimensional Dual-Tree Complex Wavelet Transform (3D DT-CWT). Directional wavelet coefficients are used to extract the wavefront set and singular support associated with metal structures, followed by morphological processing to obtain a binary metal mask. The corrupted projections are then inpainted using harmonic interpolation, and the final reconstruction is obtained by combining metal-free and metal-only reconstructions.
\emph{Main results.} The proposed method was evaluated on both simulated and clinical CBCT datasets. It consistently achieved more accurate metal segmentation and reduced artifacts compared to conventional image-domain hard-thresholding approaches, leading to improved visual quality and fewer residual streaks. Notably, the method remains effective in challenging scenarios, including complex anatomies and cases where metal objects are partially outside the reconstruction field of view.
\emph{Significance.} This work demonstrates that analytically grounded, projection-domain segmentation based on directional wavelet analysis enables effective and robust metal artifact reduction in CBCT without relying on training data. The approach offers improved interpretability and practical advantages for clinical deployment, highlighting the potential of wavefront-set–based methods for artifact reduction in tomographic imaging.

\end{abstract}

\section{Introduction}
In medical computed tomography (CT) imaging, metals appear in, for example, dental implants and orthodontic braces.
Compared to biological tissues, metallic materials attenuate and scatter X-rays more strongly and highly non-uniformly across the X-ray energy spectrum, leading to inconsistent X-ray projections. These inconsistencies introduce severe streaking and shading artifacts in reconstructed CT images \cite{De_Man_et_al_1999__IEEETransNuclSci}, \cite{boas2012ct}, \cite{gjesteby2016metal}, 
significantly degrading image quality and compromising downstream medical image analysis. Metal artifacts are often even more severe in cone-beam computed tomography (CBCT) due to greater scatter, lower tube voltage, and cone-beam effects \cite{schulze2011artefacts}.

Metal artifact reduction (MAR) aims to mitigate these detrimental effects. From a reconstruction viewpoint, MAR algorithms can be grouped into three main categories: model-based iterative reconstruction (MBIR-MAR), image-domain MAR (ID-MAR), and projection-domain MAR (PD-MAR). Notably, most MAR methods in the literature address the 2D CT setting rather than 3D CBCT \cite{boas2012ct,schulze2011artefacts}.

MBIR-MAR methods combine a physics-based observation model with image priors in an iteratively solved optimization problem. They can produce excellent results and provide a principled way to incorporate prior information, e.g., sparsity-driven approaches \cite{choi2011sparsity}, total variation with inequality constraints \cite{schiffer2018tv}, and polychromatic statistical reconstruction \cite{bismark2020reduction}. However, MBIR approaches are typically computationally expensive due to repeated reconstruction and forward projection, limiting its practical impact.

ID-MAR methods apply post-processing to reduce streaks in reconstructed images \cite{liao2019adn,zhang2018convolutional,liang2019metal,khaleghi2021metal}. Because they operate after reconstruction, lost or distorted structures may be difficult to recover with high fidelity.

PD-MAR methods act on the projection data before reconstruction by identifying metal-corrupted regions and completing the corresponding projections. Many PD-MAR approaches treat the corrupted measurements as missing data and perform data completion using neighboring information. Although interpolation-based methods are computationally efficient, interpolation errors in the sinogram can lead to secondary artifacts in the corrected reconstructions \cite{boas2012ct,meyer2010normalized}. Normalized MAR methods mitigate this issue by interpolating normalized projections using a prior image \cite{meyer2010normalized,meyer2012frequency}. After projection completion, a metal-only reconstruction can be inserted into the artifact-reduced one to obtain the final result.

PD-MAR is often viewed as consisting of two steps: (1) metal segmentation and (2) inpainting. Accurate segmentation is critical, and simple thresholding is often insufficient \cite{boas2012ct}. After segmentation, inpainting can be performed using either traditional algorithms or learning-based approaches.

Metal segmentation can be performed in either the image or sinogram domain. Image-domain segmentation is more intuitive and commonly used in practice. The most straightforward approach is thresholding to isolate high-attenuation regions, for example using adaptive thresholding \cite{niblack1986,bernsen1986,sauvola2000}. While fast and simple, thresholding is sensitive to beam hardening and artifacts and may fail with overlapping tissues or metals of varying densities. In CBCT, where Hounsfield Unit (HU) values are less reliable than in CT, thresholding can also yield inaccurate metal boundaries. Nevertheless, a common workflow is to threshold in the image domain and forward project the resulting metal mask to obtain a metal trace in the sinogram. For example, \cite{lee2020metal} uses image-domain thresholding in dental CT before applying a fully connected network to refine a linearly interpolated sinogram. Atlas-based segmentation \cite{rohlfing2004atlas} can leverage anatomical priors, but requires accurate registration and may be brittle in complex cases.

Most PD-MAR methods therefore segment metal in the image domain and forward project the mask to the sinogram. Direct segmentation in the projection domain is comparatively less explored and also challenging: simple thresholding can be ambiguous due to overlaps and path-dependent attenuation, producing non-uniform and view-dependent metal traces. Nevertheless, addressing metal segmentation (and inpainting) directly in the projection domain is appealing, since it avoids introducing artifacts during reconstruction and allows the sinogram geometry to be exploited analytically.

Deep learning-based PD-MAR methods have become popular, see, e.g., \cite{ghani2019fast,gomi2019development,lee2020metal,agrawal2021metal}. In many settings, learning-based MAR improves quantitative performance relative to classical approaches, although methods operating solely via sinogram enhancement tend to underperform image-domain approaches \cite{gjesteby2016metal}. As an example, \cite{ghani2019fast} proposes a conditional GAN for security CT, where metal is segmented by image-domain thresholding followed by morphological operations, then forward projected to generate a sinogram mask; the contaminated projections are removed from the original sinogram.

Machine learning-based segmentation methods, such as CNNs or U-Nets trained on real \cite{hegazy2019u} or synthetic \cite{agrawal2023} CBCT data, can generalize across noise levels and artifact patterns. However, they require annotated training data and careful model development, and may be less transparent than analytical approaches. Moreover, in clinical environments, reliance on machine learning can increase regulatory complexity and approval timelines, since AI solutions are often classified as high-risk medical devices and are subject to evolving regulatory requirements across jurisdictions (e.g., MDR and the EU AI Act in Europe) \cite{aboy2024navigating}. Consequently, where feasible, conventional non-learned algorithms may be advantageous.

Despite the appeal of projection-domain MAR, most PD-MAR methods still rely on image-domain segmentation followed by forward projection to identify corrupted sinogram regions. This motivates analytical methods that perform robust segmentation directly in the projection domain.

To segment metals directly from the 3D sinogram, we propose a method based on the 3D Dual-Tree Complex Wavelet Transform (3D DT-CWT), which extends the DT-CWT to three dimensions \cite{chen2012efficient,wang2007video}. The 3D DT-CWT provides approximate shift invariance and directional selectivity, which are useful for analyzing 3D sinogram volumes. While the 2D DT-CWT has previously been used within iterative CT reconstruction for MAR \cite{us2019combining}, here we use the 3D DT-CWT to computationally extract the wavefront set corresponding to metal objects in the CBCT 3D sinogram volume, followed by morphological processing to obtain a segmentation. After segmentation, inpainting is performed in the projection domain to complete the MAR pipeline. We evaluate the method on clinical scanner data, including both clinical acquisitions and simulated metals.

The main novelty of this work is the use of the 3D DT-CWT for direct analytical metal segmentation, enabling extraction of metal traces from volumetric projection data without image-domain thresholding or learned models. 3D DT-CWT provides approximate shift invariance together with strong directional selectivity in three dimensions, enabling robust extraction of metal features directly from the 3D sinogram. This is particularly beneficial in CBCT, where metal traces propagate continuously across projection angles and slices. Unlike CNN-based segmentation methods, the proposed framework is fully analytical and training-free, requiring no annotated datasets or retraining across scanners or acquisition protocols. In contrast to previous 2D DT-CWT-based MAR approaches, where the transform was incorporated within iterative reconstruction algorithms, our method applies the 3D DT-CWT directly to the full 3D CBCT sinogram volume for projection-domain metal segmentation. This enables exploitation of the volumetric geometric continuity of metal traces across projection views, which cannot be captured by independent 2D processing.

The remainder of the paper is organized as follows. Section \ref{sec:CBCT} reviews the framework of CBCT. Section \ref{sec:dtcwt} introduces the DT-CWT and describes its use for singularity extraction. Section \ref{sec:method} presents the proposed MAR method step-by-step. Section \ref{sec:comparison_methods} describes the comparison method based on image-domain hard thresholding. Section \ref{sec:datasets} describes the simulated and experimental datasets. Section \ref{sec:metrics} details the quantitative evaluation metrics, and Section \ref{sec:results} presents the results. Finally, Sections \ref{sec:discussion} and \ref{sec:conclusion} conclude with a discussion and directions for future work.

\section{Methods}
\subsection{Cone-beam computed tomography} \label{sec:CBCT}

CBCT is a lightweight and low-cost alternative to conventional CT. In CBCT, the patient is scanned using a cone-shaped beam and an X-ray detector panel for direct 3D image reconstruction. The 2D projection images of the field of view (FOV) are acquired during a single rotation. CBCT scanning cuts the radiation exposure when compared to helical CT, as the scanning FOV is smaller and tightly cropped \cite{kalender2011computed}. Common applications of CBCT are dental imaging \cite{kiljunen2015dental}, extremity imaging and orthopedic studies \cite{planmed}, and oral and maxillofacial imaging \cite{angelopoulos2012comparison}. 

The benefits of CBCT are its mechanical simplicity, mobility, low cost, and high spatial resolution compared to conventional CT. However, the cone-shaped X-ray beam, slow rotation, and limited detector area may lead to artifacts and low temporal resolution. Furthermore, the increased scattering with CBCT produces inferior soft-tissue contrast, magnified metal artifacts, and poor HU stability, which have limited the widespread applicability of CBCT \cite{kiljunen2015dental}. Metals are a significant source of artifacts in dental CBCT, which complicates clinical decision making \cite{bamberg2011metal}. To address these challenges, novel reconstruction algorithms and MAR algorithms tailored for CBCT are needed to increase diagnostic image quality \cite{mahnken2003new}, \cite{wei2004x}, \cite{bal2006metal}, \cite{prell2009novel}.

The projection data in CBCT imaging forms a 3D $xyz$-volume, also referred to as the 3D sinogram. Each $xy$-plane corresponds to a single 2D projection image, and the projection angle proceeds along the $z$-axis. Each $xz$-plane of the 3D projection data superficially resembles a conventional 2D sinogram. However, only data from the midplane, perpendicular to the axis of rotation and containing the X-ray point source, is a true sinogram in the sense of containing the Radon transform of the object in that plane \cite{Schulze_et_al_2009__ClinOralImplantsRes}.

CBCT geometry requires demanding reconstruction mathematics. The key problem with exact 3D reconstruction is that the three-dimensional sinogram does not represent a Radon space. There is a method to overcome this problem, but still, the assumption is that a complete set of Radon data is available \cite{buzug2009computed}. For the popular circular X-ray trajectory used in most technical applications, a complete set of Radon data is not available. Luckily, there are approximation methods that can also deal with incomplete Radon data. The most frequently used method for cone-beam reconstruction is the Feldkamp-Davis-Kress (FDK) algorithm \cite{feldkamp1984practical}, which is an approximate extension of the 2D filtered backprojection (FBP) algorithm into 3D cone-beam geometry. 

\subsubsection{Metal artifacts in CBCT} \label{sec:2a}
Metals in the scanned object cause inconsistencies in the projection data, resulting in artifacts in the reconstruction. Metal artifacts typically appear as bright or dark streaks surrounding metal objects in the CBCT image, as dark regions between metal objects, and as cupping effects.
The inconsistencies are due to a number of mechanisms, most importantly beam-hardening, scatter, noise, photon starvation, the non-linear partial volume effect, and aliasing \cite{De_Man_et_al_1999__IEEETransNuclSci}, \cite{boas2012ct}, \cite{gjesteby2016metal}.
The ramp filtering and backprojection in FDK reconstruction spread these inconsistencies into global artifacts. Compared to CT, metal artifacts are exacerbated in CBCT \cite{schulze2011artefacts}.

The standard reconstruction model in X-ray tomography is based on Beer-Lambert law, where the X-ray intensity registered at detector pixel $p$ is
\begin{equation}
    \label{eq:lambert_beer}
    I(p) = I_0 \, e^{-\int_\mathrm{ray\:path} \mu(x, y, z) \, ds},
\end{equation}
where $I_0$ is the intensity of incident X-ray beam and $\mu(x, y, z)$ is the distribution of attenuation coefficients in the object. However, clinical scanners use polychromatic X-ray sources, and the attenuation model must be extended to
\begin{equation}
    I(p) = \int\limits_0^{E_\mathrm{max}} I_0(E) \, e^{-\int_\mathrm{ray\:path} \mu(x, y, z, E) \, ds} \, dE.
\end{equation}
Metals attenuate different X-ray energies highly nonlinearly, and lower energy photons experience extremely strong attenuation, and the average beam energy increases, \emph{i.e.} hardens, as it passes through the medium. Furthermore, due to beam-hardening, the ratio of photoelectric absorption and Compton scattering changes for the remaining higher energy photons, with Compton effects dominating. As a result, photon paths are altered and they are registered at the detector off the center line of the incident beam. These effects causes strong inconsistencies in the observed integral attenuation values. If photon starvation occurs, practically all photons along a given ray path are absorbed and very strong streak artifacts arise.

Metals can cause a very high portion of the incident radiation to be absorbed. Due to low photon counts, the signal registered behind metals is very noisy as result of photon Poisson statistics and detector electronic noise. The random noise effects are propagated across the image during reconstruction, resulting in streaking artifacts.

The finite detector pixel size and reconstruction grid size can lead to non-linear partial volume (NLPV) effects when a metal object only partially covers a given voxel, resulting in incorrect attenuation estimation. The finite number of projection directions can also lead to streak-like artifacts around metals due to aliasing.

\subsection{3D dual-tree complex wavelet transform} \label{sec:dtcwt}

Analogous to Fourier analysis, wavelet methods decompose a function into translated and scaled copies of a mother wavelet. Wavelets have proven highly effective in signal processing applications, such as image compression. However, standard discrete wavelet transforms (e.g. Haar or Daubechies) suffer from shift sensitivity and exhibit poor directional selectivity, particularly in higher dimensions. The dual-tree complex wavelet transform \cite{kingsbury1998dual,selesnick2005dual} addresses these limitations by employing two parallel wavelet decompositions, which can be interpreted as the real and imaginary components of a complex-valued transform. This dual-tree construction yields approximate shift invariance and significantly improved directional selectivity, providing 28 distinct directions in three dimensions. 


The 3D DT-CWT consists of complex-valued scaling and wavelet functions. Consider a complex, approximately analytic wavelet \(\psi(x)\), defined by
\begin{equation}
\psi(x) = \psi_h(x) + i \psi_g(x),
\end{equation}
where \(\psi_h\) and \(\psi_g\) are the real and imaginary parts of \(\psi(x)\), respectively, and associated with a high-pass filter \(H\). Similarly, a complex scaling function \(\phi(x)\) is given by
\begin{equation}
\phi(x) = \phi_h(x) + i \phi_g(x),
\end{equation}
and associated with a low-pass filter \(L\). The real parts \(\psi_h\) and \(\phi_h\) correspond to the real components of the wavelet and scaling function, while \(\psi_g\) and \(\phi_g\) are the imaginary components. For computational purposes, we use finitely supported wavelets, yielding approximately analytic wavelets.

We define the complex wavelet coefficients \(d_\nu(j,n) \in \mathbb{C}\) by 
\begin{equation}
\begin{aligned}
d_\nu(j,n)
&= \langle f, \Psi_{\nu,j,n} \rangle \\
&= \int_{\mathbb{R}^3}
   f(x,y,z)\,
   \Psi_{\nu,j,n}(x,y,z)\,
   \mathrm{d}x\,\mathrm{d}y\,\mathrm{d}z ,
\end{aligned}
\end{equation}
where \(j = 0, \dots, J\) represents the scale and \(n = (n_1, n_2, n_3) \in \mathbb{Z}^3\) denotes the spatial index. The index \(\nu\) represents the oriented subbands, and there are 28 distinct directional subbands at each scale \(j\), indexed by:
\begin{equation}
\begin{aligned}
\mathcal{I} = \{\, &
\overline{L}HH,\;
\overline{H}LH,\;
\overline{H}HL,\;
LHH,\;
HLH,\;
HHL, \\
& HHH,\;
\overline{H}\overline{H}H,\;
\overline{H}H\overline{H},\;
\overline{L}HL,\;
\overline{L}LH,\;
\overline{H}HH,\;
\ldots
\,\}.\nonumber
\end{aligned}
\end{equation}
These 28 subbands represent the high-pass (detail) information in the transform, oriented in different directions along the three spatial dimensions. The subbands correspond to combinations of separable high-pass filters across the three axes, with each of the 7 fundamental orientations being replicated across 4 distinct quadrants (combinations of positive/negative real and imaginary parts).

The 3D wavelets are constructed as tensor products of the 1D scaling and wavelet functions. For example, the wavelet in the subband \(HHH\) is defined by:
\begin{equation}
\begin{aligned}
&\Psi_{HHH,j,n}(x,y,z) \\
= \quad & \psi(2^j x - n_1)\,
  \psi(2^j y - n_2)\,
  \psi(2^j z - n_3).
\end{aligned}
\end{equation}
Similarly, for a conjugate subband, such as \(\overline{H}HH\), we have:
\begin{equation}
\begin{aligned}
&\Psi_{\overline{H}HH,j,n}(x,y,z) \\
= \quad &\overline{\psi(2^j x - n_1)}\,
   \psi(2^j y - n_2)\,
   \psi(2^j z - n_3).
\end{aligned}
\end{equation}
In this manner, each of the 28 subbands corresponds to a distinct combination of wavelet functions along the three coordinate axes, yielding directional wavelets with high selectivity in 3D.

For discrete 3D images of size \(2^J \times 2^J \times 2^J\), the indices \(n_1, n_2, n_3\) range from \(0\) to \(2^j-1\) for each scale \(j\). The transform starts at the finest scale, applying the analysis filters to the voxel data, and proceeds toward coarser scales by filtering the low-pass subbands.

The function \(f\) can be reconstructed from its coefficients via:
\begin{equation}
\begin{aligned}
f(x,y,z)
&= \sum_{\nu \in \mathcal{I}}
   \sum_{j=1}^J
   \sum_{n_1=0}^{2^j-1}
   \sum_{n_2=0}^{2^j-1}
   \sum_{n_3=0}^{2^j-1} \\
&\qquad d_\nu(j,n)\,
        S^{-1}\Psi_{\nu,j,n}(x,y,z).
\end{aligned}
\end{equation}
where \(S = T^* T\) is the frame operator, with \(T\) being the analysis operator and \(T^*\) the synthesis operator.

\begin{figure}[!t]
    \centering
    \includegraphics[width=\linewidth]{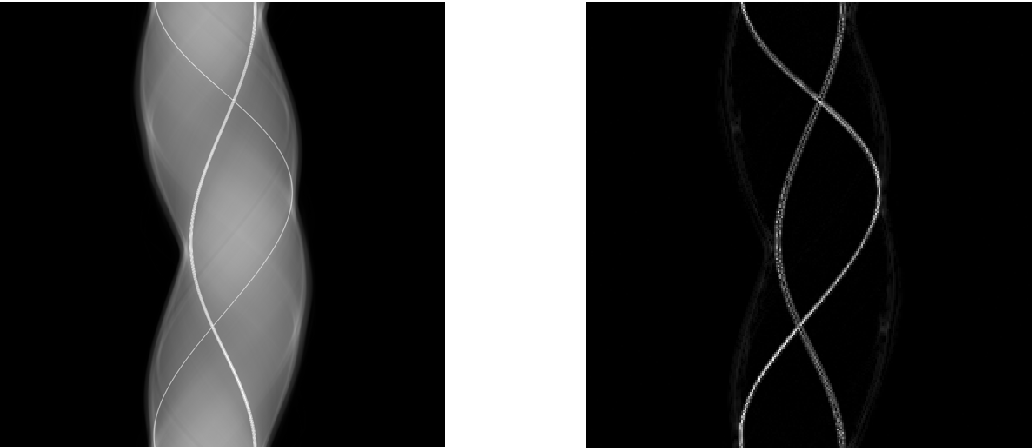}
    \caption{Sinogram (left) and the sum of the absolute value of the largest complex wavelet coefficients, revealing the location of the metals (right). }
    \label{fig:sino_cont_metals}
\end{figure}

\subsubsection{Wavelets in singularity extraction}

The wavefront set $WF(f)$ characterizes both the spatial locations and the orientations of singularities of a function. Let 
$f \in L^2_{\mathrm{loc}}(\mathbb{R}^2)$. We say that $f$ is microlocally smooth at 
$x_0$ if one can find a cutoff function $\phi \in C_c^\infty$ with $\phi(x_0)\neq 0$ such that the Fourier transform of the localized function $\phi f$ decreases faster than any polynomial in $|\xi|$ as $|\xi| \to \infty$. If no such rapid decay occurs, then $x_0$ lies in the singular support of $f$.

Directional information is obtained by examining the decay of the localized Fourier transform within angular regions of the frequency domain. If decay is not rapid inside a conical neighborhood centered at direction $\theta_0$, then $(x_0,\theta_0)$ is an element of the wavefront set $WF(f)$. In imaging problems, $\theta_0$ corresponds to the local edge orientation perpendicular to an interface.

In practice, the complex wavelet transform is an effective tool for extracting singularities from a signal, as previously demonstrated for CT wavefront set extraction in \cite{rautio2023learning}. In images and volumetric data, large-magnitude wavelet coefficients typically occur near jumps and edges. See Fig. \ref{fig:sino_cont_metals} for an example of a sinogram containing two metal inserts, where the sum of the absolute values of the wavelet coefficients reveals the corresponding metal boundaries.

The DT-CWT offers directional selectivity with moderate redundancy and computational efficiency. Real-valued wavelets provide only limited orientation discrimination, while curvelets \cite{candes2004new} and shearlets \cite{kutyniok2012shearlets,labate2005sparse} achieve finer directional resolution at increased computational cost. For large 3D volumes, the DT-CWT provides a suitable balance between directionality and computational feasibility.  

\begin{figure*}[!t]
    \centering
    \resizebox{\textwidth}{!}{\input{workflow}}
    \vspace{-1cm}
    \caption{The proposed workflow for metal artifact reduction, based on projection-domain metal segmentation. Steps include: A. Extracting the wavefront set related to metals from the 3D sinogram. B. Binary mask for metal segmentation. C. Sinogram inpainting and metal-free reconstruction. D. Metal reconstruction. E. Final result: combining the artifact-free inpainted reconstruction and metal reconstruction.}
    \label{fig:workflow}
\end{figure*}

\begin{figure*}[!t]
  \centering
  \resizebox{\textwidth}{!}{%
  \begin{tabular}{@{}r@{}c@{}c@{\hspace{0.2cm}}c@{}c}
    & \multicolumn{1}{c}{\small $z=119$} & \multicolumn{1}{c}{\small $y=155$} & \multicolumn{1}{c}{\small $z=120$} & \multicolumn{1}{c}{\small $y=335$} \\

    \raisebox{1.2cm}{3D sinogram }
    & \includegraphics[width=3.5cm]{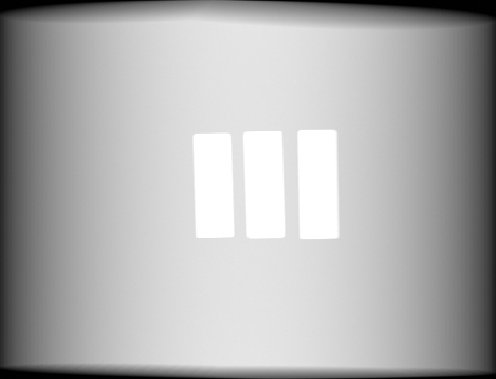}
    & \includegraphics[width=3.5cm]{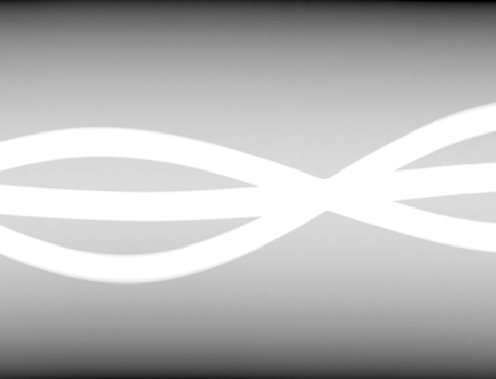}
    & \includegraphics[width=3.5cm]{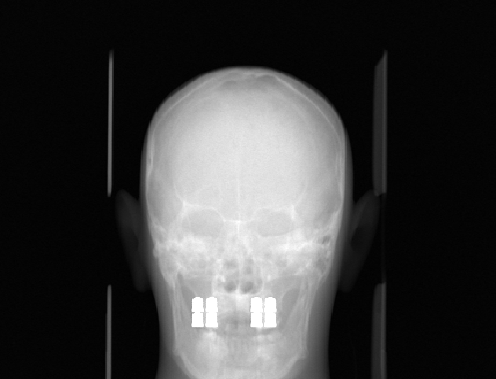} 
    & \includegraphics[width=3.5cm]{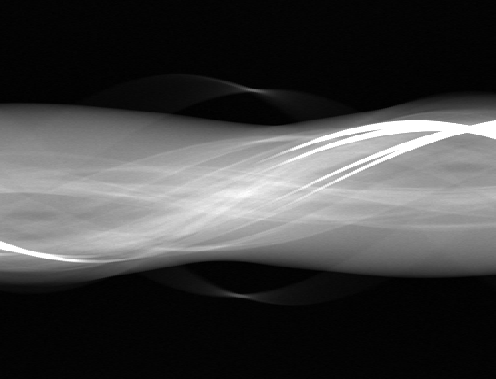} \\

    \raisebox{1.2cm}{\small $A_{\nu,j}(x,y,z)$ }
    & \includegraphics[width=3.5cm]{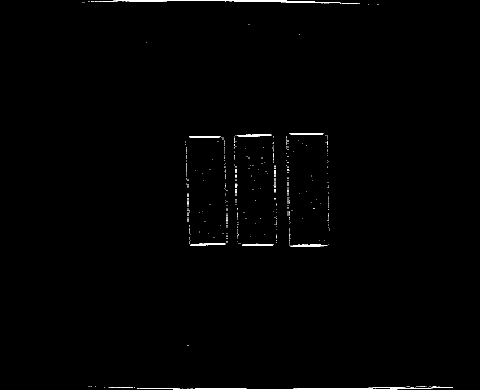}
    & \includegraphics[width=3.5cm]{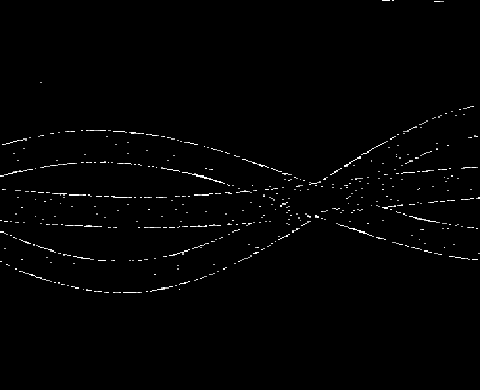}
    & \includegraphics[width=3.5cm]{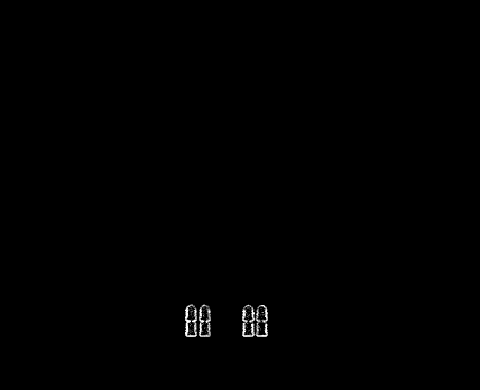}
    & \includegraphics[width=3.5cm]{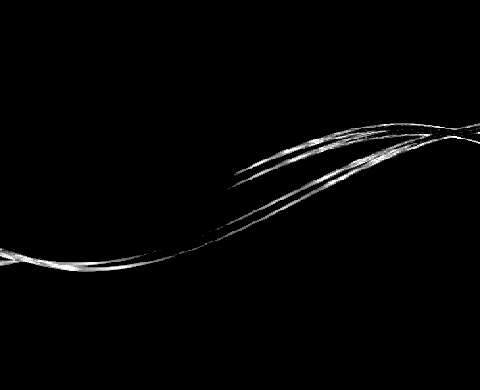} \\

    \raisebox{1.2cm}{\small metal mask }
    & \includegraphics[width=3.5cm]{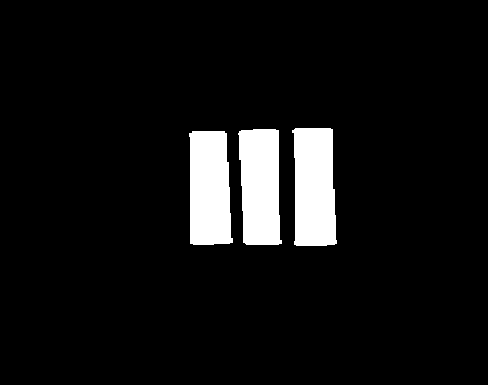}
    & \includegraphics[width=3.5cm]{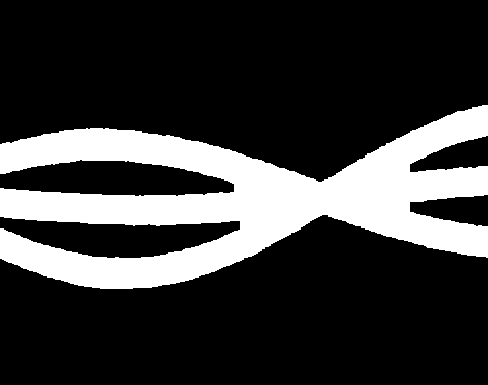}
    & \includegraphics[width=3.5cm]{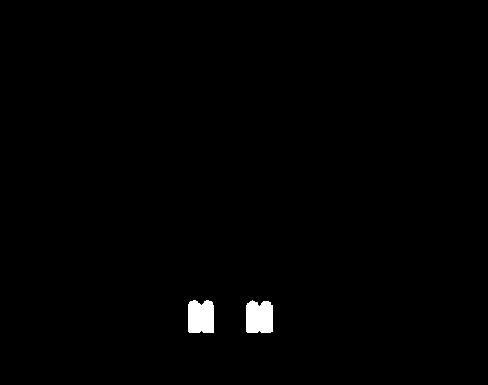}
    & \includegraphics[width=3.5cm]{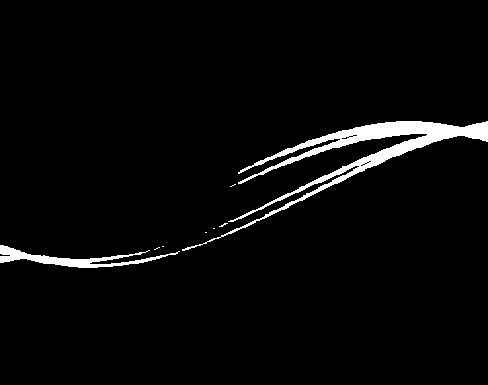} \\

    \raisebox{1.2cm}{\small metal-removed sinogram }
    & \includegraphics[width=3.5cm]{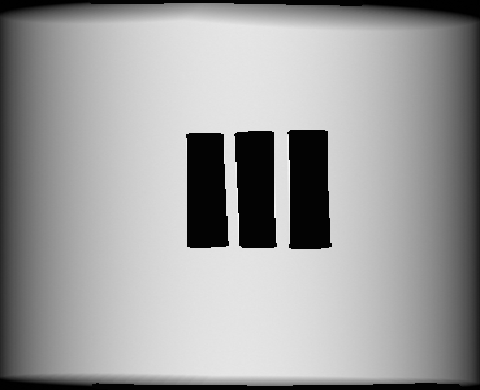}
    & \includegraphics[width=3.5cm]{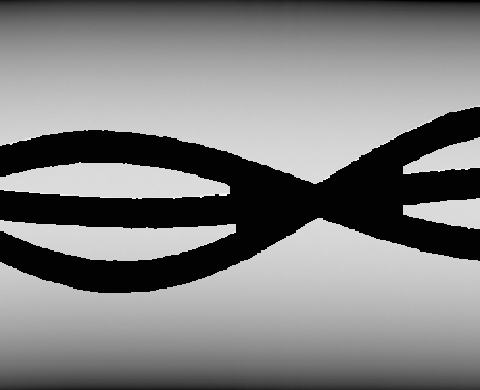}
    & \includegraphics[width=3.5cm]{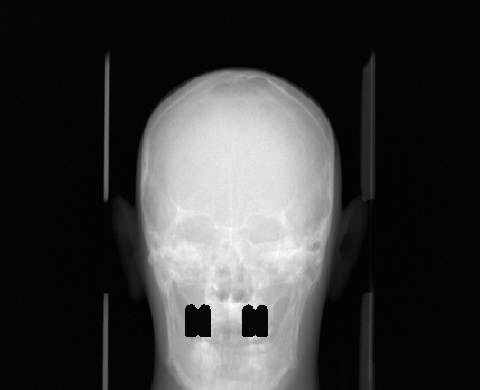}
    & \includegraphics[width=3.5cm]{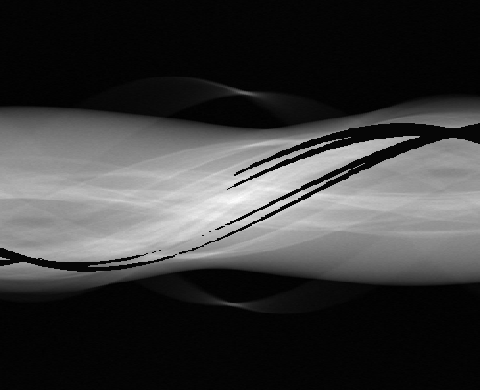} \\

    \raisebox{1.2cm}{\small metal sinogram }
    & \includegraphics[width=3.5cm]{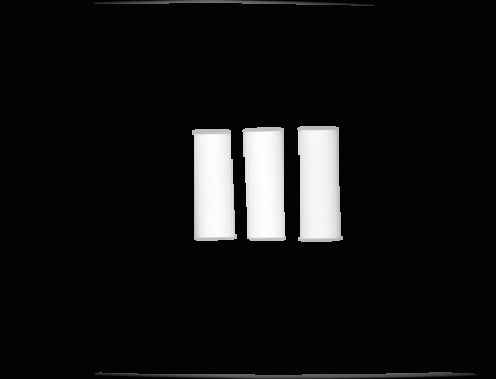}
    & \includegraphics[width=3.5cm]{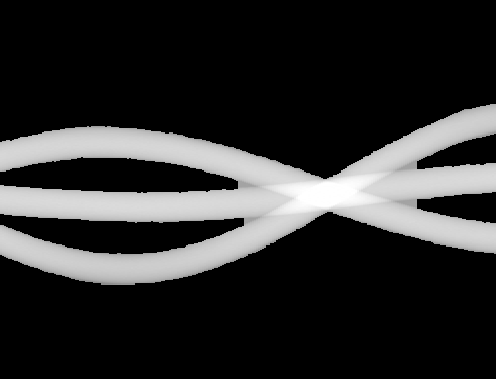}
    & \includegraphics[width=3.5cm]{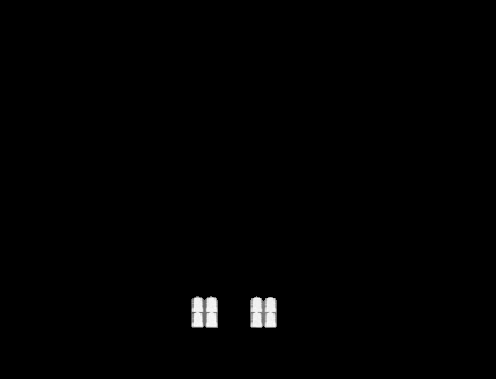}
    & \includegraphics[width=3.5cm]{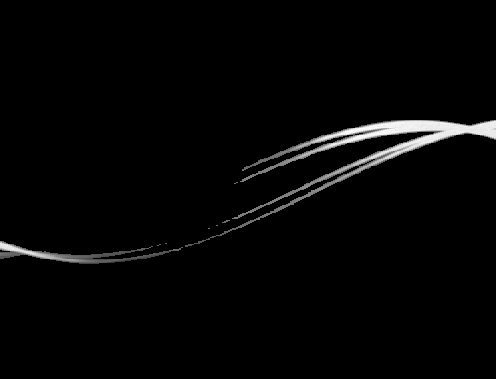} \\
  \end{tabular}
  }
  \caption{Metal segmentation results for simulated cases (I) and (II), presenting $xy$- and $xz$-slices of the 3D CBCT sinograms. Note that the metal mask slightly overshoots the metal area over the neighboring slice due to the morphological closing operation.}
  \label{fig:sinogram_segmentation}
\end{figure*}

\subsection{Proposed projection-domain MAR method} \label{sec:method}
We propose a MAR method for projection-domain
metal segmentation and inpainting, that uses 3D DT-CWT to find the edges of metals in the 3D sinogram. Fig. \ref{fig:workflow} shows the full workflow from 3D sinogram to metal artifact reduced reconstruction. The method is explained in detail below.

\subsubsection{Extracting the wavefront set related to metals}

We start with the 3D sinogram $m \in \mathbb{R}^{x \times y \times z}$ as the input, and take the DT-CW transform of it, resulting in 28 detail coefficient subbands $d_\nu(j,n) \in \mathbb{C}^{28\times n}$, where $\nu=\{1,...,28\}$ is the subband index and $n=xyz$ is the number of voxels. In this work, only the first decomposition scale, $j=1$, is used. The DT-CWT was computed using MATLAB's 'dualtree3' implementation with the default first-level biorthogonal filters and a 10-tap Hilbert Q-shift filter pair.

Metals are highly attenuating, so they correspond to  large wavelet coefficient values. We threshold away the smallest coefficients related to noise and non-metal features, by keeping only the largest coefficient values:

\[
\widetilde{d_\nu}(1,n)
=
\begin{cases}
d_\nu(1,n), & \lvert d_\nu(1,n)\rvert > \tau,\\[4pt]
0, & \lvert d_\nu(1,n)\rvert \le \tau,
\end{cases}
\]
where $\tau = 0.1$ is an empirically selected threshold applied to the normalized wavelet coefficients.

Note that coefficient thresholding might not fully eliminate non-metal contributions, since superimposed structures in the 3D sinogram may also generate singularities detected by the DT-CWT, potentially leaving residual coefficients in the subbands $\widetilde{d_\nu}(j,n)$. This will be addressed in the next step.

\subsubsection{Binary mask for metal segmentation}

To move from the complex wavelet coefficient space $\mathbb{C}^{28\times x \times y \times z}$ to the 3D sinogram space $\mathbb{R}^{x \times y \times z}$, we take a sum over the absolute value of the thresholded coefficients: 
\begin{equation}\label{abscomplexwav}
A_{\nu,j}(x,y,z) = \sum_{\nu=1}^{28}
|\widetilde{d_\nu} (1,n)|,
\end{equation} 
revealing the locations in the 3D sinogram space corresponding to the largest coefficients. See the 2nd row of Fig. \ref{fig:sinogram_segmentation} for an illustration of the resulting point cloud $A_{\nu,j}(x,y,z)$ for two different 3D sinograms. Then, the result is binarized to $B(x,y,z) \in \{0,1\}^{x \times y \times z}$ by:
\begin{equation}
B(x,y,z)
=
\mathbf{1}\!\left(A_{\nu,j}(x,y,z) > 0\right),
\end{equation}
where $\mathbf{1}$ denotes the indicator function.

As mentioned in the previous subsection, 
the binary point cloud $B(x,y,z)$ might still have voxels that are not related to a true metal trace, due to superimposition of features in the 3D sinogram. To remove these residuals, we leverage the geometric structure of sinograms and discard isolated voxels that do not align along a continuous geometric trajectory in the $z$-direction. That is, a voxel is retained only if it has nonzero support within an $n_h$-neighbourhood in the $(x,y)$ plane in at least $d$ neighbouring slices along the $z$-direction, within a prescribed depth $d$:
\begin{equation}
\begin{aligned}
& C(x,y,z) \\
= \quad &B(x,y,z)\,
  \mathbf{1}\!\Big(
  \exists\, t \in \{1,\dots,d\} \text{ such that } \quad\\
  &\sum_{(x',y') \in \mathcal{N}_{n_h}(x,y)} 
  B(x',y',z+t) +
  B(x',y',z-t)
  > 0
  \Big).
\end{aligned}
\end{equation}
with neighbourhood
\[ \mathcal{N}_{n_h}(x,y)
=
\{(x',y') : |x'-x|\le n_h,\ |y'-y|\le n_h\}.\]
In experiments (I-II, IV-V), we used a neighbourhood of $n_h =5$ and depth $d=4$ to preserve continuous metal trajectories. In experiment (III), $(n_h,d)=(6,8)$ was used to account for a weaker continuity of the metal trace.

Voxels that fail this connectivity test are removed as isolated noise. The connectivity-filtering procedure is applied iteratively until convergence, i.e., until no further voxels are removed between successive iterations. By iterating this procedure, the remaining structures exhibit coherent propagation across slices, resembling the flow of information in a CBCT sinogram. The resulting values $C(x,y,z) \in \{0,1\}^{x \times y \times z}$ form a point cloud that corresponds to the singular support of the metal traces in the 3D sinogram space.

Next, we convert $C(x,y,z)$, which contains points located on metal boundaries, into a full binary mask for metal segmentation. Since the extracted boundary may not form a closed surface, we complete it using standard morphological operations \cite{gonzalez2018digital}. In particular, we apply morphological closing, which consists of a dilation followed by an erosion using the same structuring element $S \subset \mathbb{Z}^3$. This operation fills small gaps and discontinuities in the extracted boundary while preserving the overall shape of the metal objects.

The morphological closing of a set $\mathcal{D}$ by a structuring element $S$ is defined as
\begin{equation}
\mathcal{D} \bullet S = (\mathcal{D} \oplus S) \ominus S,
\end{equation}
where $\oplus$ and $\ominus$ denote morphological dilation and erosion, respectively. In our implementation, we apply morphological closing to $C(x,y,z)$ using a spherical structuring element $b$ of radius $3$, yielding
\begin{equation}
\widetilde{C}(x,y,z) = C(x,y,z) \bullet b_{r=3}.
\end{equation}
The morphological closing operation is applied iteratively until the extracted boundary forms a closed surface.

After the boundary has been closed, the interior is filled using a flood-fill operation (MATLAB's 'imfill' function). The resulting binary mask $M_{\text{metal}} \in \{0,1\}^{x \times y \times z}$ represents the segmented metal regions in the 3D sinogram (see the third row of Fig. \ref{fig:sinogram_segmentation}).

The computational cost for the proposed metal segmentation is as follows. For the largest test case, dataset (IV) with 3D sinogram of size $858 \times 858 \times 500$, the complete segmentation pipeline including DT-CWT computation, coefficient thresholding, connectivity filtering, and morphological post-processing, required 81.8s and reached a peak memory consumption of 26.7GB. Runtime measurements were performed on a workstation equipped with an Intel Xeon Gold 6128 CPU (6 cores, 12 threads) and 53GB RAM.

\subsubsection{Sinogram inpainting and metal-free reconstruction}

Using the metal mask \(M_{\text{metal}}\), voxels corresponding to metal structures are segmented from the original 3D sinogram \(m\), yielding a metal-removed sinogram \(m_{\text{NM}}\). See the fourth row of Fig. \ref{fig:sinogram_segmentation}. The missing regions are filled using harmonic inpainting by solving the discrete Laplace equation within the masked region subject to Dirichlet boundary conditions prescribed by the surrounding known pixel values using MATLAB’s 'regionfill' function. 

Each \(xy\)-projection of the 3D sinogram is inpainted independently. Inside the metal mask \(M_{\mathrm{metal}}\), the inpainted values are defined as the solution of the discrete harmonic equation
\begin{equation}
m_{\mathrm{inp}}(x,y,z)
= \frac{1}{4}\!\!\sum_{(i,j)\in\mathcal{N}(x,y)} m_{\mathrm{inp}}(i,j,z),
\end{equation}
where $(x,y)\in M_{\mathrm{metal}}(\cdot,\cdot,z)$, and \(\mathcal{N}(x,y)\) denotes the four-connected neighborhood in the \(xy\)-plane. Values outside the mask are kept fixed to the original projection, imposing Dirichlet boundary conditions.

After the metals have been inpainted from the 3D sinogram, a reconstruction is computed using the FDK algorithm. The reconstruction $f_{\text{inp}}$ does not include metals and, as such, should have reduced metal artifacts.

\subsubsection{Metal reconstruction}
For the final result, the metal components need to be added back into the inpainted reconstruction \(f_{\text{inp}}\). To this end, the metals are reconstructed from a metal-only 3D sinogram obtained via metal segmentation, using the mask $M_{\text{metal}}$. The resulting metal FDK reconstruction \(f_{\text{M}}\) contains artifacts due to the inaccurate monochromatic approximation of the polychromatic Beer--Lambert law, as discussed in Section~\ref{sec:2a}. However, since \(f_{\text{M}}\) consists solely of metal objects on an otherwise empty background, these artifacts can be readily removed by thresholding, thereby retaining only the metal information.

\subsubsection{Final result}
Finally, the metal reconstruction \(f_{\text{M}}\) is added to the metal-free reconstruction \(f_{\text{inp}}\), yielding an artifact-free reconstruction with the metal objects restored.

\subsection{Comparison methods} \label{sec:comparison_methods}
We benchmark the proposed method against metal segmentation based on hard thresholding (HT) in the image domain. This comparison method is referred to as HT-MAR throughout the paper. The threshold is selected in the range 2300--3000~HU, depending on the dataset. Following thresholding, the segmentation is morphologically dilated using a spherical structuring element \( b_{r=3} \) to slightly expand the metal regions and better capture boundary voxels. The resulting segmentation is then forward-projected to obtain the corresponding metal mask in the projection domain for inpainting. All subsequent steps follow the same pipeline as in the proposed method.

\begin{figure*}[!t]
  \centering
  \resizebox{\textwidth}{!}{%
  \begin{tabular}{@{}r@{}c@{\hspace{0.1cm}}c@{\hspace{0.1cm}}c@{\hspace{0.1cm}}c}
    & \multicolumn{1}{c}{\small projection} & \multicolumn{1}{c}{\small PD-HT} & \multicolumn{1}{c}{\small Otsu} & \multicolumn{1}{c}{\small CW-MAR} \\

    \raisebox{1.2cm}{\small $z=3$ }
    & \includegraphics[height=3.5cm]{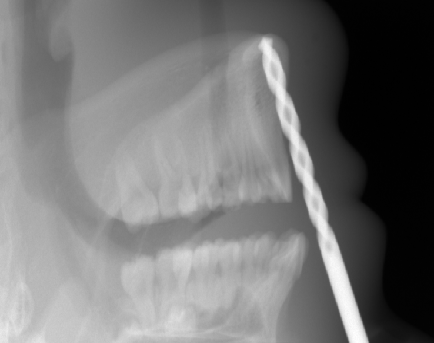}
    & \includegraphics[height=3.5cm]{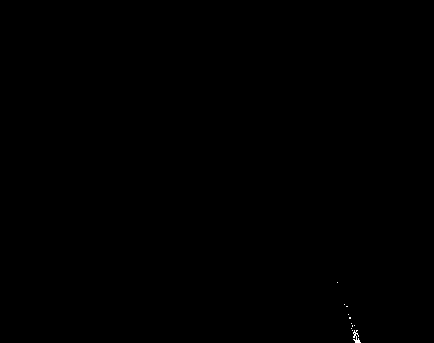}
    & \includegraphics[height=3.5cm]{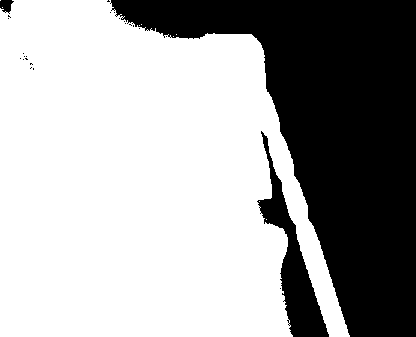}
    & \includegraphics[height=3.5cm]{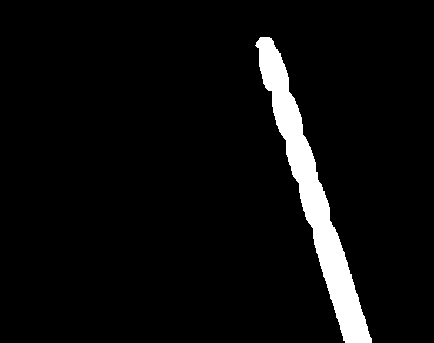} \\

    \raisebox{1.2cm}{\small $z=190$ }
    & \includegraphics[height=3.5cm]{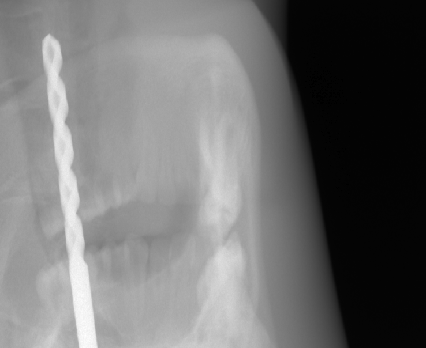}
    & \includegraphics[height=3.5cm]{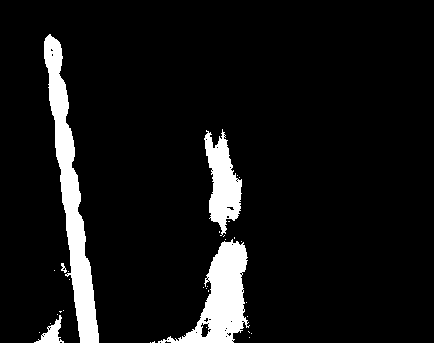}
    & \includegraphics[height=3.5cm]{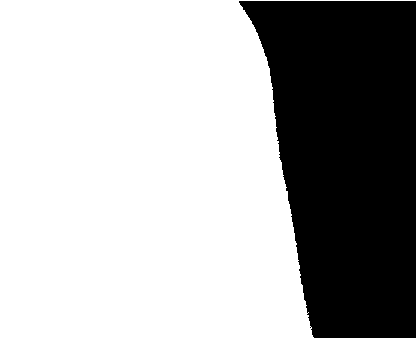} 
    & \includegraphics[height=3.5cm]{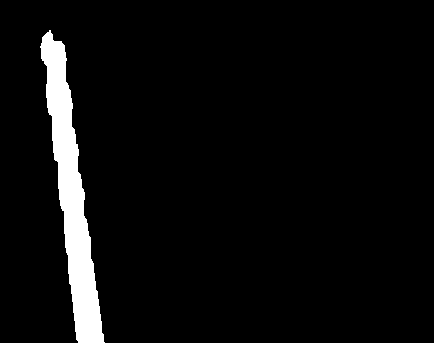} \\
 
  \end{tabular}
  }
    \caption{Comparing different projection-domain metal segmentation techniques for experimental phantom (V): projection-domain hard-thresholding, Otsu's method, and proposed CW-MAR. The other methods fail to detect the metal trace, whereas the proposed CW-MAR method successfully recovers its overall shape, enabling subsequent projection-domain inpainting. }
    \label{fig:PDHT}
\end{figure*}

Additionally, we evaluated hard thresholding (PD-HT) and Otsu's thresholding \cite{otsu1979threshold} directly in the projection domain. See Figure \ref{fig:PDHT} for comparisons against the proposed CW-MAR for projection domain slices from the experimental phantom (V). The two projection-domain comparison methods do not detect the metal trace due to the superposition of structures in the projections which causes non-metal regions to exhibit high voxel values while, conversely, some metal features may not be distinctly visible. As a result, direct voxel-value based projection-domain thresholding fails to provide a reliable metal segmentation. As both approaches prove ineffective, they are therefore excluded from further analysis.

\subsection{Datasets} \label{sec:datasets}
We evaluate the proposed method using physical phantoms with both simulated and experimental metal inserts. The datasets with simulated metals include: (I) a PMMA phantom containing three simulated titanium inserts, and (II) an anthropomorphic phantom containing eight simulated amalgam inserts. In both cases, metal artifacts were simulated using a Monte Carlo–based X-ray transport simulator following \cite{agrawal2024deep,agrawal2025}. Specifically, we inserted metals in CT volumes and simulated primary and scattered projections along with flat-field projections for the geometry of Viso G7 CBCT device.

The experimental datasets include: (III) a tooth embedded in gelatin with an amalgam filling, (IV) a high-gold alloy fixed dental prosthesis, and (V) an anthropomorphic phantom containing a metal screw partially outside the reconstruction field of view.

All measurements were acquired using a Planmeca Viso G7 CBCT system. Measurements (I), (II), and (V) were performed at Planmeca Group, Helsinki, Finland, while measurements (III) and (IV) were obtained from \cite{jayakody2025hybrid}, performed at the Medical Imaging Teaching and Test Laboratory (Mittlab), University of Oulu, Finland.

\subsection{Quantitative Evaluation metrics} \label{sec:metrics}

\subsubsection{Metal segmentation}
The precision of metal segmentation in the projection domain is evaluated in 3D using voxel-wise overlap between the predicted segmentation mask and the corresponding ground truth. We compute the Dice similarity coefficient,
$$D(A,B) = \frac{2 \left| A \cap B \right|}{\left| A \right| + \left| B \right|},$$
and the Jaccard similarity index,
$$J(A,B) = \frac{\left| A \cap B \right|}{\left| A \cup B \right|},$$
where \(A\) denotes the predicted segmentation and \(B\) the ground truth. Segmentation metrics are computed only for datasets (I--II), for which ground-truth segmentations are available. For the comparison method (image-domain HT-MAR), the image-domain segmentation is forward-projected and subsequently binarized prior to metric evaluation.

\subsubsection{Reconstruction quality}

Quantitative image quality was evaluated using region of interest (ROI) based artifact metrics adapted from \cite{peters2025hybrid}. Due to the absence of metal-free reference images, we employed no-reference residual streak variation and local intensity standard deviation metrics to quantitatively compare the quality of reconstructions.

Residual streak variation was used to quantify the magnitude of local intensity oscillations caused by streak artifacts. For a region containing a prominent streak artifact, the metric was defined as
\begin{equation} \label{eq:streak}
V_{\mathrm{streak}} = P_{95}(I_{\mathrm{ROI}})-P_{5}(I_{\mathrm{ROI}}),
\end{equation}
where $P_{95}$ and $P_{5}$ denote the 95th and 5th percentiles of the voxel intensities within the ROI, respectively. Lower values of the metric indicate reduced residual streak artifacts.

In addition, we compute the local intensity standard deviation, defined here as
\begin{equation} \label{eq:stdev}
\sigma_{\mathrm{LI}} =
\sqrt{
\frac{1}{N-1}
\sum_{i=1}^{N}
\left(I_i-\bar{I}\right)^2
},
\end{equation}
where $I_i$ denotes the intensity of voxel $i$, $\bar{I}$ is the mean intensity within the ROI, and $N$ is the number of voxels in the ROI. Lower standard deviation values indicate a more homogeneous region and fewer residual artifacts.

\begin{figure*}[h!]
\centering
\begin{picture}(550,550) 
\put(50,543){Uncorrected FDK}
\put(210,543){HT-MAR}
\put(345,543){CW-MAR}
\put(16,417){\includegraphics[width=15cm]{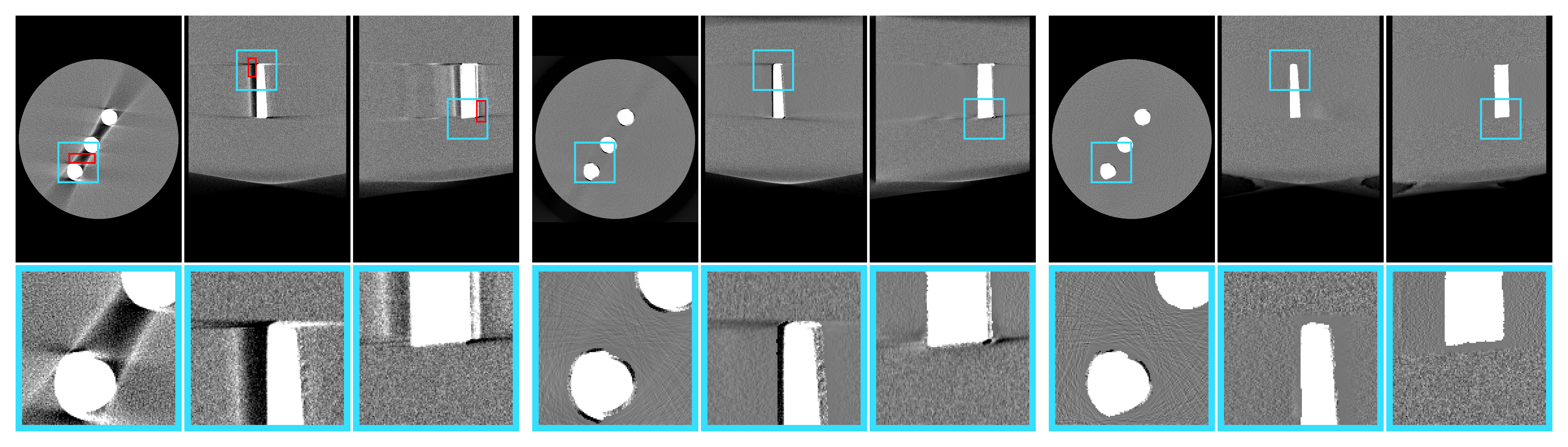}}
\put(7,467){(I)}

\put(16,293){\includegraphics[width=15cm]{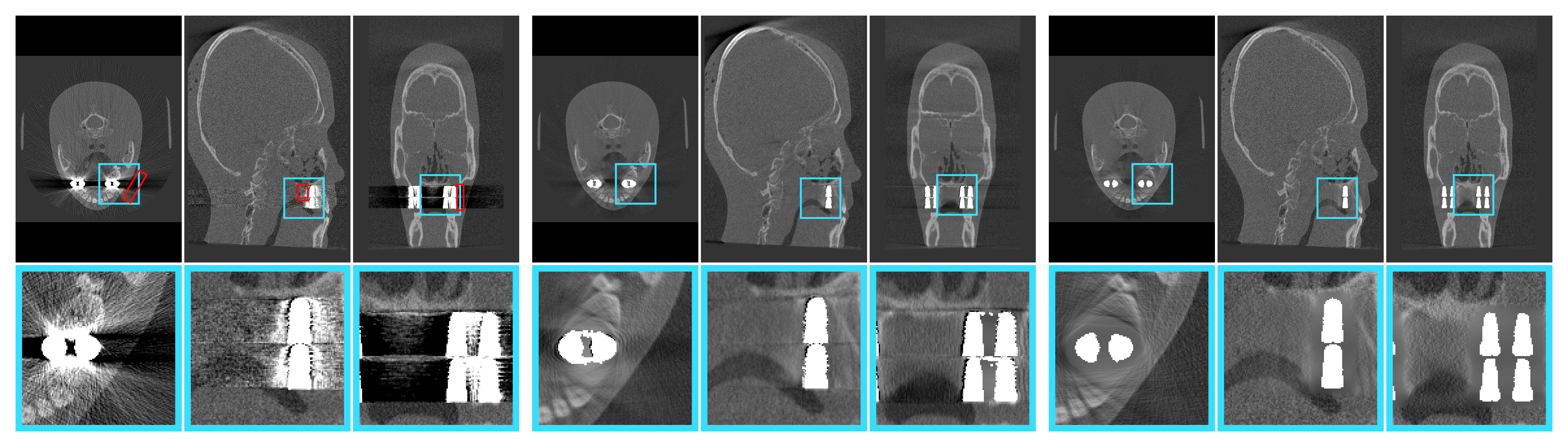}}
\put(3,340){(II)}

\put(16,204){\includegraphics[width=15cm]{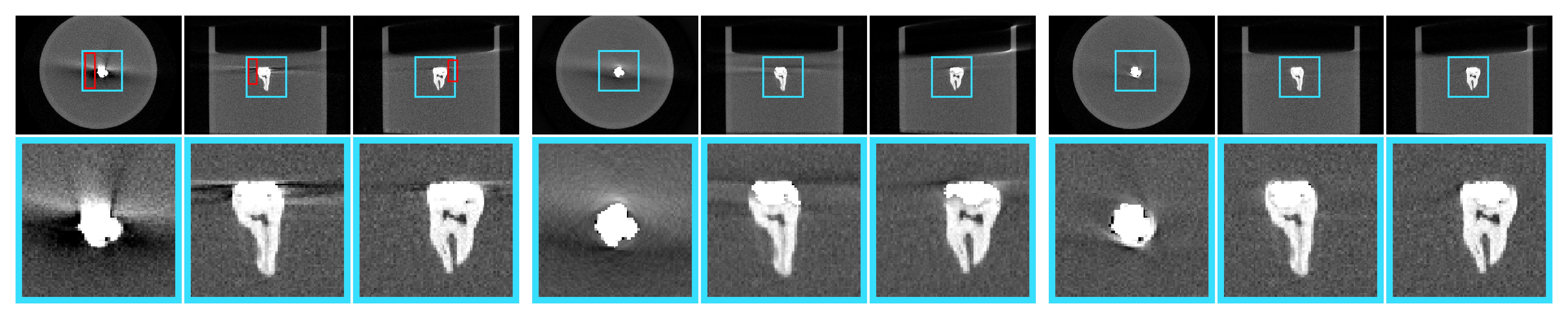}}
\put(0,250){(III)}

\put(16,102){\includegraphics[width=15cm]{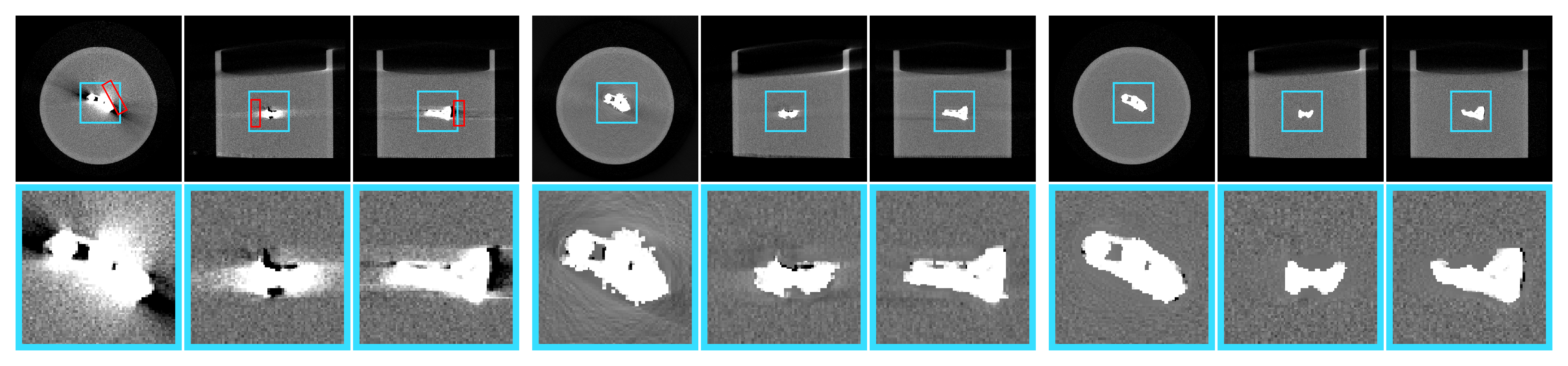}}
\put(0,147){(IV)}

\put(16,0){\includegraphics[width=15cm]{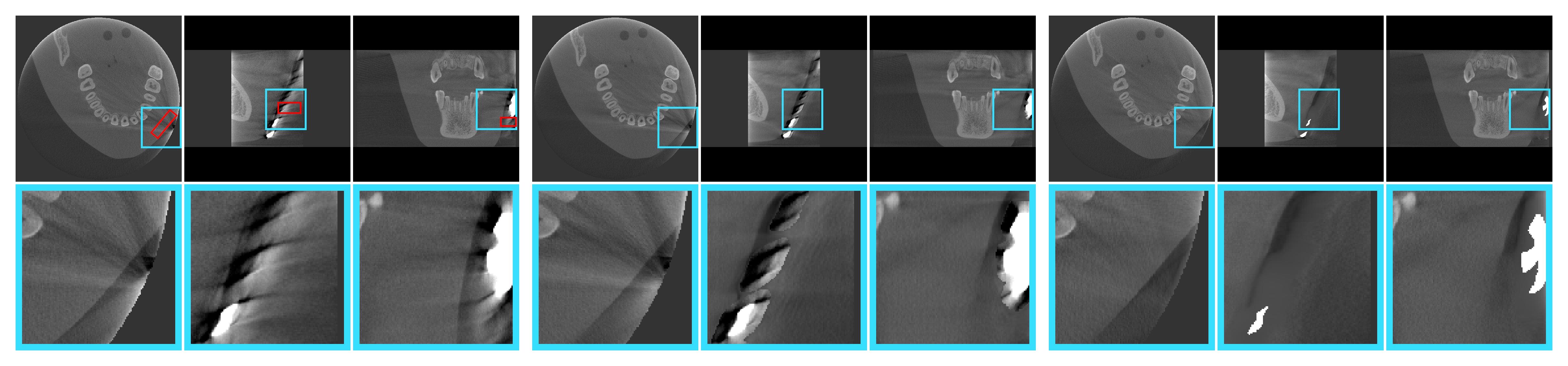}}
\put(3,48){(V)}

\end{picture}
\caption{Comparison of FDK reconstructions before correction and after artifact reduction using HT-MAR (comparison method) and CW-MAR (proposed method). Results are shown for physical phantoms with simulated metals (I–II) and experimental metals (III–V). For each phantom and method, axial, sagittal, and coronal slices are displayed. The red boxes on the FDK reconstructions indicate the representative ROIs used for the quantitative evaluation. Reconstructions are not HU-calibrated. Visualization uses dataset-specific HU-equivalent windows to highlight metal-induced artifacts. Windowing (range, center): (I) \([-1000, +1400]\)~HU (200~HU), (II) \([-2700, +5700]\)~HU (1500~HU), (III) \([-996, +186]\)~HU (\(-405\)~HU), (IV) \([-998, 0]\)~HU (\(-654\)~HU), and (V) \([-1240, -30]\)~HU (\(-635\)~HU).
}
\label{fig:all_results}
\end{figure*}

For each dataset (I--V), ROIs were manually selected in regions exhibiting visible streak artifacts while avoiding the metal objects themselves and strong anatomical or phantom boundaries. ROIs were selected on representative axial, coronal, and sagittal slices, and the reported values were obtained by averaging the metric over the selected directions. The same ROI locations were used for both metrics and all compared reconstruction methods, see the red areas indicated in Figure \ref{fig:all_results}. See Table \ref{tab:metrics} for the quantitative results.

\section{Results} \label{sec:results}
We computed the uncorrected FDK reconstruction, the image-domain hard-thresholding--based MAR reconstruction, and the proposed projection-domain complex wavelet MAR reconstruction for datasets (I--V). Fig. \ref{fig:all_results} shows selected slices from the reconstruction volumes for each dataset, displayed using identical windowing. The quantitative results of the metal segmentation are reported in Table~\ref{tab:sinogram_metrics} for datasets (I) and (II).

Dataset (I) contains three simulated titanium inserts in an otherwise homogeneous PMMA phantom. Consequently, metal segmentation in both the image and projection domain is expected to be relatively straightforward. As seen in Fig.\ref{fig:all_results}, the uncorrected FDK reconstruction exhibits substantial metal artifacts, which are particularly prominent in the axial slices. The HT-MAR method is able to reduce these artifacts to a large extent. However, residual artifacts remain near the metal boundaries as a result of imperfect segmentation. Even after morphological widening of the image-domain segmentation, the mask remains too narrow in some regions while being overly wide in others. This behavior is evidenced by the close-up reconstruction images and the segmentation metrics reported in Table \ref{tab:sinogram_metrics}. In contrast, the proposed CW-MAR method yields a more accurate projection-domain segmentation, resulting in an almost complete suppression of metal artifacts due to the tighter and more consistent segmentation. This leads to the lowest residual streak variation and to local intensity standard deviation in Table \ref{tab:metrics}.

Dataset (II) presents a considerably more challenging MAR problem, featuring eight metal implants in a more complex anthropomorphic phantom. The uncorrected FDK reconstruction is again heavily affected by metal artifacts, which are prevalent in all slicing directions and obscure many anatomical structures. The HT-MAR method is able to reduce these artifacts to a reasonable extent, with projection inpainting restoring much of the missing information. However, the estimated metal boundaries are somewhat inaccurate and include portions of bright artifact regions. The CW-MAR method captures the metal boundaries more robustly. Nevertheless, some room for improvement remains, as projection inpainting introduces mild blurring in the vicinity of the metal regions. This is also reflected in the numerical metrics: while CW-MAR results in the lowest values, there is still some streak variation present in the HU values. 

First of the experimental metal datasets, dataset (III) provides a simple test case for severe artifacts caused by an amalgam filling. As in dataset (I), the HT-MAR approach reduces artifacts in the uncorrected FDK reconstruction. However, residual artifacts persist due to missed metal voxels near the metal boundaries in the segmentation. The CW-MAR method removes nearly all artifacts, particularly in the sagittal and coronal slices. The reduction of streaking artifacts is quantitatively supported in the metrics presented in Table \ref{tab:metrics}.

Dataset (IV) produces substantial artifacts in the FDK reconstruction. For HT-MAR, identifying a suitable threshold level is challenging, as the HU values of the artifacts overlap with those of the actual metal features. This leads to poor metal segmentation. In contrast, CW-MAR removes most artifacts and recovers the general shape of the prosthesis. The quantitative metrics also support these findings.

Dataset (V) presents a particularly challenging MAR scenario, in which the metal object is partially outside the reconstruction field of view (FOV). As a result, streaking artifacts are present even when the metal itself does not appear within the reconstructed volume. The large screw produces severe artifacts in the uncorrected FDK reconstruction. The image-domain HT-MAR approach is unable to reduce artifacts caused by metals outside the reconstruction FOV and additionally struggles with accurate boundary segmentation, especially in the coronal direction, where the screw geometry is particularly challenging. In contrast, the CW-MAR method is not affected by metals outside the reconstruction FOV, since segmentation is performed in the 3D sinogram domain, where the corresponding metal trace is present. Consequently, artifact reduction is effective. The quantitative metrics report largely reduced artifacts for the CW-MAR method, while values for FDK and HT-MAR remain similar, indicating a failure of artifact removal. However, as in previous cases, projection inpainting introduces some blurring in CW-MAR, which in this dataset leads to a loss of fine anatomical detail near the metal boundary.

\begin{table}[t]
\centering
\begin{tabular}{c l c c}
\hline
\textbf{Dataset}& \textbf{Metric} & \textbf{HT-MAR} & \textbf{CW-MAR} \\
\hline
(I)&Dice  & 0.8375 & \textbf{0.8876}\\
&Jaccard  & 0.7204 & \textbf{0.7980}\\
\hline
(II)&Dice  & 0.8523 & \textbf{0.8562}\\
&Jaccard  & 0.7426 & \textbf{0.7486}\\
\hline
\end{tabular}
\caption{Quantitative comparison of metal segmentation performance for the HT-MAR and proposed CW-MAR methods on datasets (I) and (II), using ground-truth masks derived from simulated metal inserts. The best-performing result for each metric is shown in bold.}
\label{tab:sinogram_metrics}
\end{table}

\begin{table*}[h!]
\centering
\begin{tabular}{lccc}
\toprule
\textbf{Dataset} & \textbf{Method} & \textbf{Streak} & \textbf{StDev} \\
\midrule
 & FDK & 1655.2 ± 108.9 & 518.4 ± 58.4  \\
\textbf{(I)} & HT-MAR & 697.5 ± 446.9 & 244.3 ± 100.7  \\
& CW-MAR  & $\bold{401.8 \pm 28.9}$ & $\bold{124.1 \pm 10.3}$ \\
\midrule
&FDK & 2892.0 ± 878.8 & 919.1 ± 282.4 \\
\textbf{(II)}&HT-MAR & 743.0 ± 138.1 & 245.9 ± 61.1 \\
&CW-MAR  & $\bold{649.9\pm158.7}$ & $\bold{193.7\pm50.0}$ \\
\midrule
&FDK & 329.4 ± 244.4 & 102.4 ± 75.8  \\
\textbf{(III)}&HT-MAR & 169.1 ± 77.5 & 54.6 ± 28.2  \\
&CW-MAR  & $\bold{120.2\pm10.2}$ & $\bold{36.6\pm5.0}$ \\
\midrule
&FDK & 634.9 ± 146.4 & 198.7 ± 48.8  \\
\textbf{(IV)}&HT-MAR & 153.4 ± 29.3 & 61.5 ± 27.5  \\
&CW-MAR  & $\bold{119.6\pm12.5}$ & $\bold{37.9\pm1.4}$  \\
\midrule
&FDK & 453.4 ± 313.8 & 139.9 ± 96.4  \\
\textbf{(V)}&HT-MAR & 413.9 ± 317.5 & 127.2 ± 81.1  \\
&CW-MAR  & $\bold{73.3\pm9.4}$ & $\bold{23.2\pm1.9}$ \\
\bottomrule
\end{tabular}
\caption{Quantitative results assessing the reduction of metal artifacts. 'Streak' refers to the residual streak variation (Eq. \ref{eq:streak}), and 'StDev' to the local intensity standard deviation (Eq. \ref{eq:stdev}). Lower values indicate successful artifact removal. The best-performing result for each metric is shown in bold.}
\label{tab:metrics}
\end{table*}

\section{Discussion} \label{sec:discussion}

This study investigated projection-domain metal segmentation in 3D cone-beam computed tomography using the dual-tree complex wavelet transform. By exploiting directional wavelet coefficients to extract the wavefront set of metal-induced singularities, the proposed method enables segmentation directly in the 3D sinogram domain without relying on image-domain thresholding or training data. The results demonstrate that this approach yields accurate and geometrically consistent metal masks, which in turn, enable effective projection-domain inpainting and artifact reduction.

Direct thresholding in the sinogram domain is unreliable in CBCT due to the superposition of structures and view-dependent attenuation, which motivates the use of directional multiscale analysis for projection-domain metal segmentation. These results further highlight the advantage of the proposed DT-CWT-based segmentation approach in complex 3D sinogram data.

Compared to conventional image-domain hard-thresholding methods, the proposed approach provides more robust segmentation of metal traces, particularly in challenging scenarios where attenuation values are unreliable or where structures overlap in projection data. This improved segmentation leads to a more complete suppression of streak artifacts and fewer residual inconsistencies in the reconstructed volumes. Notably, the method remains effective when metal objects lie partially outside the reconstruction field of view, a scenario in which image-domain approaches typically fail due to missing information in the reconstructed volume. This highlights a key advantage of operating directly in the projection domain.

The use of the 3D DT-CWT provides a computationally efficient means of capturing directional information in volumetric sinogram data, offering a practical balance between directional selectivity and computational cost. 
Alternative directional multiscale transforms, such as shearlets and curvelets, also provide strong directional selectivity and could be explored for projection-domain metal segmentation. However, these transforms typically involve higher redundancy and computational cost than the DT-CWT, particularly for large 3D volumes. In this work, we selected the DT-CWT because it provides a practical balance between directional selectivity, computational efficiency, and implementation complexity.

We did not include deep learning-based MAR methods in the experimental comparison, since the primary aim of this work is to investigate an analytical, training-free projection-domain segmentation framework. While learning-based approaches can achieve strong performance, they typically require large annotated datasets and retraining across acquisition settings, whereas the proposed method is scanner-agnostic, interpretable, and does not rely on training data.

\subsection{Limitations and Future Work}

The reconstruction results presented in this work employ harmonic inpainting, which may introduce local blurring near metal regions. Since harmonic inpainting reconstructs missing projection data by enforcing smoothness, fine structures or high-frequency information that are completely obscured by the metal trace cannot be fully recovered. As a result, anatomical details adjacent to large metallic objects may be partially smoothed, particularly in anatomically complex regions where the missing data span a wide area of the sinogram. However, the proposed DT-CWT-based metal segmentation framework is not restricted to harmonic inpainting and can be combined with alternative projection-domain inpainting techniques.

Although the proposed method improves metal segmentation accuracy, residual non-metal structures may still be included in highly complex projection data due to superposition effects. Furthermore, the current implementation requires empirically selected parameters, including the wavelet threshold and connectivity-filtering parameters, which may require adjustment for different scanners, anatomies, or metal types.

Another limitation of the proposed method is the memory-intensive nature of the 3D DT-CWT. Although the method is suitable for offline CBCT processing, the large volumetric datasets and multiple directional subbands can lead to substantial memory requirements.

Future work will focus on improving reconstruction quality near metal boundaries by incorporating more advanced projection-domain inpainting strategies or hybrid approaches that combine analytical segmentation with data-driven refinement. Developing fully automated parameter selection strategies that generalize across different scanners, anatomies, and metal types is another important direction for future research. More memory-efficient implementations, alternative multiscale transforms, and more sophisticated morphological processing may further improve segmentation accuracy while reducing computational cost. Despite these limitations, the proposed method demonstrates that analytically grounded, projection-domain segmentation is a viable and effective approach for metal artifact reduction in 3D CBCT imaging.

\section{Conclusion} \label{sec:conclusion}

We proposed a projection-domain metal artifact reduction method for 3D cone-beam computed tomography based on wavefront-set–guided segmentation using the three-dimensional dual-tree complex wavelet transform. By performing metal segmentation directly in the 3D sinogram, the method avoids limitations associated with image-domain approaches and enables effective inpainting and artifact reduction. Experimental results on simulated and clinical datasets demonstrate improved segmentation accuracy and substantial reduction of metal artifacts compared to conventional hard-thresholding methods.

The proposed approach provides a non-learned, interpretable, and computationally efficient alternative to data-driven methods, with particular advantages in challenging scenarios such as out-of-field-of-view metals. These findings highlight the potential of analytical, geometry-aware methods for robust artifact reduction in tomographic imaging.

\section*{Acknowledgment}
This work was supported by Business Finland (decision no. 8132/31/2022), and by the Research Council of Finland through the Flagship of Advanced Mathematics for Sensing, Imaging and Modelling (grant no. 359182 and 359186) and the Centre of Excellence in Inverse Modelling and Imaging (grant no. 353097).

The authors would like to thank Annina Sipola and Ritva Näpänkangas from University of Oulu for arranging the prosthetic materials for datasets (III) and (IV).

\bibliographystyle{dcu}
\bibliography{bib}

\end{document}